\documentclass[conference,compsoc]{IEEEtran}

\usepackage{tikz}
\usepackage{amsmath}

\usepackage{filecontents}
\usepackage[linesnumbered,algoruled,boxed,lined, noend]{algorithm2e}
\usepackage{algpseudocode}
\usepackage{array}
\usepackage{verbatim}
\usepackage{mdwmath}
\usepackage{eqparbox}
\usepackage{stfloats}
\usepackage{url}
\usepackage{xcolor}
\usepackage{makecell}
\usepackage{CJK} 
\usepackage{multirow}

\usepackage{hyperref}
\usepackage[all]{hypcap}
\usepackage{etoolbox}
\usepackage{booktabs}
\usepackage[most]{tcolorbox}
\usepackage{tabularx}
\usepackage{hyperref}

\usepackage[hyphenbreaks]{breakurl}
\usepackage{CJK}
\usepackage[russian,greek,main=english]{babel}
\usepackage{authblk}

\newcommand{\tech}{\mbox{\textsc{BanAdv}}}

\makeatletter
\g@addto@macro{\@algocf@init}{\SetKwInOut{Parameter}{Parameters}}
\makeatother

\begin{document}

\title{\Large \bf Obfuscating IoT Device Scanning Activity via Adversarial Example Generation}

\author[1]{Haocong Li}
\author[2]{Yaxin Zhang}
\author[3]{Long Cheng}
\author[1]{Wenjia Niu}
\author[4]{Haining Wang}
\author[1]{Qiang Li}

\affil[1]{Beijing Jiaotong University}
\affil[2]{China CITIC Bank}
\affil[3]{Clemson University}
\affil[4]{Virginia Tech}

\renewcommand\Authands{ and }

\maketitle

\begin{abstract}

Nowadays, attackers target Internet of Things (IoT) devices for security exploitation, and search engines for devices and services compromise user privacy, including IP addresses, open ports, device types, vendors, and products.
Typically, application banners are used to recognize IoT device profiles during network measurement and reconnaissance.
In this paper, we propose a novel approach to obfuscating IoT device banners ({\tech}) based on adversarial examples. 
The key idea is to explore the susceptibility of fingerprinting techniques to a slight perturbation of an IoT device banner. By modifying device banners, {\tech} disrupts the collection of IoT device profiles. To validate the efficacy of {\tech}, we conduct a set of experiments. 
Our evaluation results show that adversarial examples can spoof state-of-the-art fingerprinting techniques, including learning- and matching-based approaches.
We further provide a detailed analysis of the weakness of learning-based/matching-based fingerprints to carefully crafted samples. Overall, the innovations of {\tech} lie in three aspects: (1) it utilizes an IoT-related semantic space and a visual similarity space to locate available manipulating perturbations of IoT banners; (2) it achieves at least 80\% success rate for spoofing IoT scanning techniques; and (3) it is the first to utilize adversarial examples of IoT banners in network measurement and reconnaissance.

\end{abstract}

\section{Introduction}

Nowadays, a wide range of IoT devices, including surveillance cameras, routers, and home automation systems, have become an integral part of our daily lives, and there will be an estimated 41.6 billion connected IoT devices worldwide by 2025~\cite{idc-report}. As IoT devices become visible and accessible on the Internet, they have emerged as prime targets for adversaries, due to their expansive attack surface, including weak credentials, insecure protocols, and inadequate software security.
Cyberattacks on IoT devices have led to severe consequences, ranging from DDoS attacks~\cite{antonakakis2017understanding}, compromising local networks~\cite{iot-attack-1}, breaking into homes~\cite{fernandes2016security}, and disrupting critical infrastructures~\cite{gaslline}.

IoT device scanning campaigns have attracted significant interest from the academic and industry community.
Prior works~\cite{formby2016s, fachkha2017internet, nmap, kumar2019all, song2020cam} have leveraged application banners and other host information to recognize IoT device profiles during the network measurement and reconnaissance.
Both attackers and search engines use application banners to achieve objectives.
From an attack perspective, IoT device scanning is the first stage of an intrusion attempt, allowing an attacker to locate, enumerate, and target vulnerable devices.
Many search engines (Shodan~\cite{Shodan}, Censys~\cite{Censys}, Zoomeye~\cite{Zoomeye}, FoFa~\cite{FoFA} and BinaryEdge~\cite{BinEdge}) offers detailed device profiles that compromise user privacy, including IP addresses, open ports, services, device types, vendors and products.

Despite much attention to IoT in the security community~\cite{fernandes2016flowfence, he2018rethinking, jia2017contexlot, wang2018fear}, little has been done to thwart or interfere IoT device scanning activities. 
In this paper, we present a novel approach to interfering with IoT device scanning activity ({\tech}) by generating adversarial examples, which slightly change IoT device banners.
Recent research shows that the performance of machine learning algorithms degrades when attackers craft variations of images~\cite{szegedy2013intriguing} and texts~\cite{papernot2016crafting}.
We leverage such an evasion technique as an efficient and preemptive approach for nullifying unsolicited IoT device scanning activities. 
Our main idea is to generate adversarial examples of IoT device banners to mislead IoT device recognition techniques, bringing about two advantages.
(1) Thwarting IoT scanning is a proactive defense strategy against potential cyberattacks, disrupting the collection of IoT device profiles.
(2) Adversarial banners can prevent the unsolicited revelation of IoT device information, protecting user privacy.

There are two significant challenges to overcome. The first technical challenge is that the IoT scanning activity is a complete black box, and we have no idea what model an attacker will use or which tool they will use to identify IoT devices.
The second technical challenge is that the random perturbation in device banners may impede normal usage of IoT devices.
Existing adversarial techniques~\cite{goodfellow2014explaining, moosavi2016deepfool, carlini2017towards, papernot2016limitations, wang2021adversarial} are not designed for nullifying IoT device recognition techniques. 
For instance, some IoT banners, such as login and configuration web pages, can be rendered to the user interface.
Randomly changing the IoT device banner can result in visual inconsistencies for users and a different appearance from the original.

To address these technical challenges,
{\tech} pursues two directions for finding an available perturbation of the IoT banners: a visual similarity space and an IoT-related semantic space. 
The visual similarity space uses Unicode specifications~\cite{unicode}, where a subtle perturbation goes unnoticed by the human's eyes but misleads IoT device scanners.
The IoT-related semantic space is an IoT-related word corpus containing keywords from device banners. 
Leveraging these two spaces, we propose two evasion strategies to generate adversarial examples against learning-based and matching-based scanning techniques.
To validate its efficacy, we systematically evaluate {\tech} for generating adversarial examples of IoT device banners.
Our evaluation results show that {\tech} successfully uses crafted device banners to spoof a learning-based scanner, covering a diverse set of machine learning models, and a matching-based scanner, covering Nmap~\cite{nmap} and Zmap~\cite{durumeric2013zmap} tools. 
Compared to the baseline's performance, our adversarial banners can maintain visually consistent appearances from the user's perspective.
Further, we provide insights into the weaknesses of the learning-based/matching-based scanners to the adversarial examples.

\begin{figure*}[!t]
  \centering
  \includegraphics[width= 4.7in]{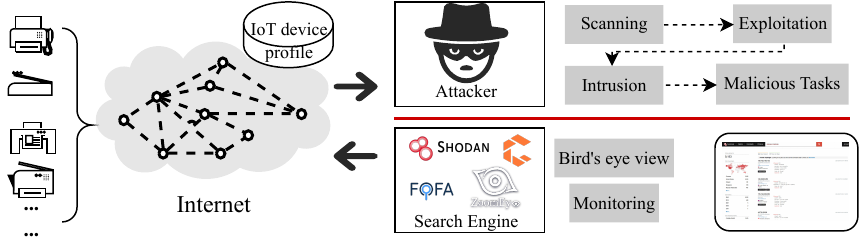}
  \caption{Network measurement and reconnaissance: attacker’s and search engine's scanning activities.}
  \label{fig:back:attack}
\end{figure*}

The major contributions of this work are summarized as follows:
\begin{itemize}
\item We propose a novel approach to misleading IoT device recognition techniques. {\tech}~\footnote{Artifact and Code: \url{https://github.com/Cong805/Spoof-IoT}} generates adversarial banners against learning-based and matching-based tools as a feasible and promising proactive defense and privacy-preserved mechanism. 
\item {\tech} achieves at least 80\% success rate for spoofing IoT scanning techniques with a 0.05 change rate, outperforming the baselines. Adversarial banners are visually consistent with original IoT device banners.
\item  We shed light on the generalization property of adversarial examples at the device banner, where the evasion technique can easily deceive IoT device scanning techniques.
\end{itemize}

\noindent
\textbf{Roadmap.}
The remainder of this paper is organized as follows.
Section 2 provides the background of IoT device scanning.
Section 3 presents the design of {\tech} against IoT device scanning technique, and Section 4 details two perturbation spaces for the {\tech}. 
Section 5 describes the experimental evaluation.
Section 6 illustrates the discussion and expansion of future work.
Section 7 surveys related work, and finally, Section 8 concludes.

\section{Background of IoT Device Scanning}

Network scanning activities have been continuously emerging.
Remarkably, these activities can successfully profile millions of devices within a few hours.
IoT device scanning comprises two primary stages: (1) probing the public address space across a spectrum of critical ports (such as 80 and 23) and application protocols (e.g., HTTP and FTP) and (2) gathering application banners to extract device information.
An IoT device scanner employs port scanning to detect active hosts and dissects handshakes to extract banners about each host and protocol. For example, application handshakes yield banners containing HTTP responses in the case of HTTP.

\textbf{Attacker's Scanning Activities}.
As a growing security concern, IoT device scanning plays a vital role in the initial stage of cyberattacks to discover and compromise vulnerable devices. A typical cyberattack follows a well-defined cycle with distinct phases characterized by network reconnaissance, exploitation, intrusive attempts, and malicious task execution.  
As the initial stage of a cyberattack, network reconnaissance involves device scanning activities on the Internet.
Networking infrastructures are 
vulnerable to malicious scanning probes, and there has been a rise in malicious scanning campaigns for IoT devices~\cite{Hershel2014, formby2016s, fachkha2017internet, miettinen2017iot, xuan2018are, kumar2019all}.

Attackers target available and visible IoT devices on a subset of IP addresses, as shown in Figure~\ref{fig:back:attack}.
Attackers are capable of sending any probing packets to those IP addresses, recording and analyzing their response packets. 
In the collection phase, the attacker uses the scanning tools (e.g., Nmap~\cite{nmap} and Zmap~\cite{durumeric2013zmap}) to find active ports for each IP address and collect its banners for each IP address, where each banner is used to construct IoT device profile.
During the profiling phase, the attacker leverages the recognition methods or tools to determine whether a banner comes from an IoT device.
To find a vulnerable target, if a banner is classified as IoT, the attacker extracts its relevant information from the banner.
Typically, for a vulnerability, a common platform enumeration (CPE)~\cite{cpe} is used to describe its 
affected product. 
An IoT device's vendor and product names are highly related to its vulnerability information.
For instance, the CVE-2023-32619~\cite{cve2023-32619} has a CPE string `cpe:2.3:o:tp-link:archer\_c50\_v3\_firmware:$\dots$:$*$', where its vendor is `tp-link' and its product is `archer\_c50'. 
Those devices with the profile of `Tp-link Archer C5v' have the vulnerability of using the hard-coded credentials as denoted by CVE-2023-32619.

\textbf{Search Engine's Scanning Activities}.
Search engines usually leverage network scanning to periodically collect the online presence of IoT devices and services, as shown in Figure~\ref{fig:back:attack}.
There are several popular search engines for devices, such as Shodan~\cite{Shodan}, Censys~\cite{Censys}, Zoomeye~\cite{Zoomeye}, FoFa~\cite{FoFA} and BinaryEdge~\cite{BinEdge}.
Those search engines offer IoT profiles about each device and service, including their IP addresses, open ports, device types, vendors, and product versions. 
The advantages of search engines are as follows: (1) help administrators protect systems against attacks, and (2) help vendors monitor their brands' online presence.

However, the view of the IoT device profile inevitably compromises user privacy.
Certain users who care about their sensitive IoT device usage are well motivated to obfuscate device banner information.  
From a privacy perspective, such scanning may cause user privacy concerns, as it reveals what devices users have and their purposes.
Data minimization and anonymization are two techniques that prevent sensitive device profiles from being collected and stored in the network measurement.
Those techniques rely on the cooperation of network administrators, which is not always feasible.


\begin{figure}[!t]
  \centering
  \includegraphics[width= 2.9 in]{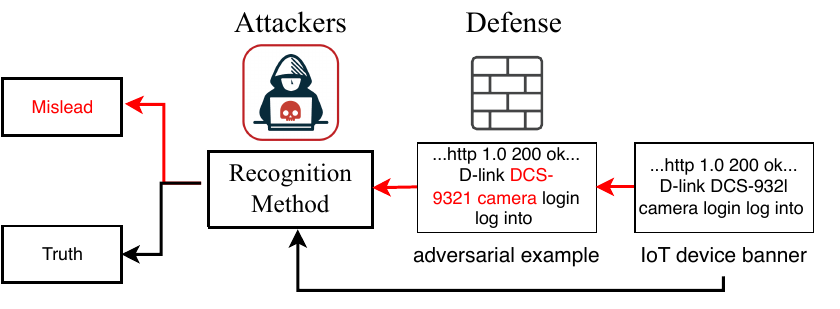}
  \caption{Adversarial examples of IoT device banners in the IoT device scanning activity.}
  \label{fig:back}
\end{figure}

\section{{\tech}: Design}
\label{sec:design}

\textbf{{\tech}'s Goal} is to generate adversarial banners against the IoT device scanning technique.
We craft an adversarial example $x'$ from a legitimate sample $x$ to spoof the mapping function $F(\cdot)$, as the following problem:
\begin{equation}\label{equ:perb}
  \begin{aligned}
    & x'    =    x + \eta, \quad F(x') \neq F(x); \\
    & sim(x, x')  \leq \epsilon
  \end{aligned}
\end{equation}
where $\eta$ is a slight perturbation added into the input's packet, and $sim(x, x')$ is the perceptual similarity between adversarial and the original examples.
As the IoT device scanner has multi-label outputs, we focus on non-targeted attacks, aiming to misclassify an IoT device label to another label to disrupt the collection of IoT device profiles. 
As shown in Figure~\ref{fig:back}, the original banner `d-link DCS-932l' is for a camera product from the D-Link vendor, and we carefully crafted an adversarial banner `d-link DCS-9321' to deceive the attacker, leading to erroneous predictions.

There are two requirements that serve as the design considerations for generating adversarial banners:
\begin{itemize}
    \item \textbf{Effectiveness}.
    {\tech} effectively spoofs IoT device scanning techniques, leading to incorrect prediction results. 
    This limitation hampers an attacker's ability to gather IoT-related information and achieve primary reconnaissance results from remote hosts. 
    \item \textbf{Minimum impact to regular usage}.
    {\tech} adds a slight perturbation in the application banner, which cannot affect the regular usage of IoT devices. 
    The adversarial banners are visual consistency for keeping the input's original appearance.    
\end{itemize}

\begin{figure}[!t]
    \centering
    \includegraphics[width=2.9in]{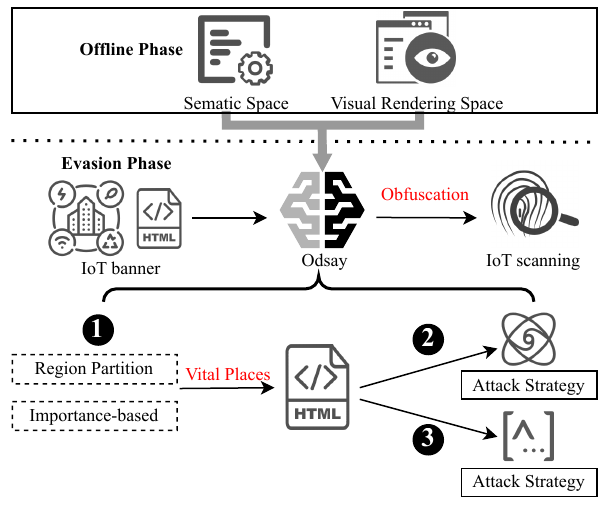}
    \caption{The architecture of Our {\tech}}
    \label{fig:arch}
\end{figure}

\textbf{Overview}.
Figure~\ref{fig:arch} presents the architecture of {\tech}.
In the offline phase, we obtain two perturbation spaces for disturbing IoT device banners, including IoT semantic and visual similarity spaces (see Section~\ref{sec:space}).
In the online phase, we first find important positions for IoT banners where we can manipulate operations. 
Different regions within the banner have varying perceptual impacts on human eyes.
To ensure regular usage, we divide these locations into three regions: immutable, focus, and non-focus zones.
After that, {\tech} proposes two evasion strategies for IoT device scanning techniques.
For the learning-based scanning technique, {\tech} generates an adversarial banner in the black-box setting in the evasion phase.
For the matching-based scanning technique, {\tech} uses a shadow model to generate an adversarial banner in the evasion phase.
Below, we present those three modules in detail.

\subsection{Finding Important Position}

To spoof IoT device scanning techniques, {\tech} locates the position/place to add a perturbation in the device banner.
Similar to the NLP task~\cite{guo2021gradient, abusnaina2020dfd, wang2022semattack}, various positions in the textual plain have different impacts on IoT device scanners. 
Note that attackers do not provide the IoT technique to defenders where the loss gradient function, model architecture, and parameters are unknown.

\textbf{Semantic Position}.
In the black-box setting, the score-based approach sends the query to the model and receives its confidential score.
Specifically, we use the score-based approach~\cite{gao2018black, jin2020bert} to calculate the confidential score of every word and find important places in the IoT device banner.
We leverage the following equation:
\begin{equation}\label{equ:imp}
    S(w_i, y)=  F_y(x) - F_y(x_{w_i\leftarrow <unk>})
  \end{equation}
where $w_i$ is the position of the word, $<unk>$ indicates to remove $w_i$, $F$ is the target model, $x$ is the original sample, and $y$ is the label. 
Comparing the prediction before and after removing a word reflects how the word influences the classification result.
Given a banner $x$, we convert it into a sequence of words, denoted as \{${w_1}, {w_2}, \ldots {w_n}$\}.
For each word $w_i$, we would calculate its important score in the IoT device banner.
Then, we rank those words in descending order by the score, with the top ones prioritizing adding a perturbation to the IoT device banner.

\textbf{Visual Position}. 
An adversarial example should be visible consistency to human eyes, affecting the IoT device scanning technique. 
Different regions in IoT banners may have various requirements for adding perturbations, which guarantee the consistency of visual information. 
We divide the IoT banner into three different regions, as follows.

(1) Immutable region (IMR). 
We cannot add any perturbation to the mutable region.
For example, the HTTP banner contains element tags, CSS styles, and Javascript in HTML content, which are used to determine the page structure and style for visual rendering effects. 
If a word in the IoT banner belongs to this region ($w_i \in IMR$),  {\tech} skips it. 

(2) Focus region (FR). 
The elements in the FR are the most visible parts of the IoT banner, easily noticed by end-users. 
For example, the banner's title and highlighted textual plain are visually significant to the users. 
The FR usually contains information important to the user, where we cannot use the semantic perturbation of the IoT to change words in this area, which requires a high degree of visual information retention.
If a word in an IoT banner belongs to this region ($w_i \in FR$), its substitute set is the Unicode character, which includes a set of visual similarity characters, zero-width characters, and reversed order characters (detailed in Section~\ref{sec:sub:unicode}).

(3) Non-focus region (NFR). 
The element in the NFR has no particular style and no highlighted areas that are not the user's focus. 
The rendering system often displays these elements as small or long text fields.
For example, words in the comment in CSS and words in paragraphs surrounded by $<$p$>$ in HTML documents.
Manipulations from the IoT-related semantic space can perturb these elements. This manipulation guarantees the semantic coherence of the banner without necessitating visual consistency for users.
If a word in the IoT banner belongs to this region ($w_i \in NFR$), its substitution set is the similarity words from the similarity space (detailed in Section~\ref{sec:space}).

The partition of the visual region relies on the application protocols of the IoT device. Specifically, the HTTP banner has three regions, while the FTP banner has two regions. This discrepancy arises because the IMR is not pertinent in the case of the FTP banner.
We determine the region of the element in the IoT banner through keyword matching.
Take an example. The region partition of the HTTP banner is shown in Table~\ref{tab:region} in the Appendix. 
When the same word appears in different regions, its disturbance operation varies.

\subsection{Evasion against Learning-based Scanner}

We use a learning-based scanner to represent the IoT detection method.
A Learning-based scanner leverages a machine learning algorithm to map an application banner into an IoT device label.
The input is the banner, denoted as $x$, and the output is the device label ($y$).
The mapping relationship can be represented as the function $F_{\theta}(\cdot)$. We directly use $F_{\theta}(\cdot)$ as the target model to generate adversarial examples of IoT device banners.

\begin{algorithm}[!t] \small
    \SetKwData{Left}{left}
    \SetKwData{This}{this}
    \SetKwData{Up}{up}
    \SetKwFunction{Union}{Union}
    \SetKwFunction{FindCompress}{FindCompress}
    \SetKwInOut{Input}{Input}
    \SetKwInOut{Output}{Output}
    \Input{Targeted model $F$, original $x$, label $y$;  \\
           Substitution set $C$;  Iteration $T$; \\
           Similarity Threshold $\theta$, Parameter $K$ }
    \Output{Adversarial example $x_{adv}$}
    \BlankLine
    $y \leftarrow F(x)$; \quad $x'\leftarrow x$;  \Comment{ $x$ is $\{w_1, w_2, \dots w_n\}$}  \\
    $W_{K}   \leftarrow  Sort(\{w_1, w_2, \dots w_n\}$)  \textit{by Equ.~\ref{equ:imp}}  \Comment{Pick top $K$}\\
    \For{$(i, w_i) \in W_{K}$}{
    
        \For{$c\in C(w)$}{                            
           $L^i$.append($x'_{w_i\leftarrow c}$);  \\ 
       }
    }
    \For{$t \leftarrow 1$ to $T$}{
        $L^{k, t} \leftarrow \arg\max$($S(x^k)$, $S(L^{k, t-1})$);   \textit{by Equ.~\ref{equ:adv:score}} \\  
        $G \leftarrow \arg\max(S(L))$;                  \\              
        \If{$F(G) \neq y $ $\&$ $sim(x, G) \leq \theta$ }
        {
            $x_{adv} \leftarrow G$; \\
            break;                \\
        }
        $x^k \leftarrow$ disturb $x^k$ by $P_L$, $P_G$ \\
        $\delta \leftarrow $ change rate($x^k, x$); \\
        $P_R = \min(0, 1-2\delta )$ \\
        $x^k \leftarrow$ random($x^k$,$P_R$); \\
    }
    \Return $x_{adv}$
    \caption{ \textit{{\tech}}: Generating adversarial examples against a learning-based scanner. }
    \label{alg:model:generated}
\end{algorithm}

\textbf{Selecting K perturbation positions}. 
Given an IoT banner, its input is $x={w_1,...,w_i,...,w_n}$ with a label $y$. 
We leverage the score-based approach (Equation~\ref{equ:imp}) to sort the $x$ and pick the top $K$ positions to add perturbations.

\textbf{Multi-Space Usage}.
We use the semantic and visual similarity spaces of the IoT to obtain the substitution set for the possible perturbation.
Given one word $w_i$, its substitution set is denoted as $C(w_i)$, where we pick $P$ candidate substitutions from the $C(w_i)$ to generate $P$ samples.
Therefore, we can obtain $K*P$ perturbation samples $\{x^1,...,x^{(1*P)},...,x^{(k *P)},...,x^{(K*P)}\}$ of the original banner to find a qualified adversarial example.

\textbf{Iteration}.
{\tech} uses the threshold $T$ as the iteration number to generate adversarial examples.
In each iteration, we would find the highest adversarial score of the population set \{$x^1$, $\dots$, $x^{(K*P)}$\}. 
For simplicity, we use the set $L={L^1,..., L^i,. ..,L^K}$ to represent $K*P$ perturbation samples, where each $L^i$ is the set \{$x^{i*1}, \dots, x^{i*P}$\}.
In each iteration, we need to find the largest influence of the perturbation of the set $L$ to improve the effectiveness of the evasion strategy.
We use the formula to find the optimal adversarial sample $G$ in the iteration, as follows:
\begin{equation}\label{equ:adv:score}
    \begin{split}
    S(x^k)  & =  F_y(x) - F_y(x^k)  \\
    L^k     & = argmax(S(x^k ),S(L^{(k,t-1)} )) \\
    G       & = (\arg\max)_{L^k\in\{L^1,...,L^k \}}(S(L^k))
    \end{split}
\end{equation}
where $F(\cdot)$ is the confidence score returned by the targeted model, $t$ is the current iteration, the $L^{(k, t-1)}$ is the optimal adversarial example in the previous iteration $t-1$.
$S(\cdot)$ is the difference between the perturbation and the original input of the targeted model.
The higher the score $S(\cdot)$, the greater the impact on the targeted model.

Once the optimal disturbed sample $G$ is obtained, we determine whether it can deceive the IoT device scanner, misleading the different label $y$.
We further use the Jaccard similarity to determine whether the  $G$ is similar to the original $x$, denoted as the $Jaccard(x, G)$. 
If the similarity is less than the threshold $\theta$, we use $G$ as an adversarial example of the IoT banner $x$.

\textbf{Mutation}.
In every iteration, we leverage two probabilities ($P_L$ and $P_G$) as the perturbation mutation when optimal sample $G$ does not meet the requirement of the adversarial example. 
We use the formula to calculate two probabilities as follows,
\begin{equation}\label{equ:update}
    \begin{split}
    P_L=(1-t/T)\times a+t/T\times b  \\
    P_G=t/T\times a+(1-t/T)\times b 
    \end{split}
\end{equation}
where $t$ is the current iteration, $a$ and $b$ are two constants, satisfying $0<b<a<1$, and $a+b=1$.
The $P_L$ refers to one perturbation operation in the original example, and the $P_G$ is the overall optimal adversarial sample from the set $L$.
Note that $P_G$ ensures the algorithm's global optimal value and $P_L$ makes the algorithm quickly converge.

To expand the disturbance mutation, we assign a random factor to the IoT banner, which is calculated by the formula as follows, 
\begin{equation}\label{equ:update3}
    P_R=\min(0, 1-2\delta) 
\end{equation}
where the $\delta$ refers to the change rate between the adversarial example and the original input,  $\delta = |x^k/x|$.
During the iteration, $P_R$ would make us pick an unmodified word that differs from the $K$ words selected by Equation~\ref{equ:imp}.

Algorithm~\ref{alg:model:generated} depicts the overall process, which generates an adversarial example in the black-box setting against a learning-based scanner.

\subsection{Evasion against Matching-based Scanner}

We use a matching-based scanner to represent the IoT detection method.
A learning-based scanner leverages string matching to map the banner $x$ into the device label $y$. 
The IoT-related keywords rely on professional researchers who find and collect those strings. 
This labor-intensive manual approach enumerates the keywords required and helps identify an IoT device.
Fortunately, many prior works~\cite{nmap, durumeric2015search, Shodan, xuan2018are} have provided IoT-related keywords to the public.
Typically, the matching-based scanner can use those keywords to obtain the mapping function $F(x)=y$, where $F(\cdot)$ denotes the form \{string/regex $\rightarrow$ label\}.

However, the regex set is not a classification model, and we cannot use it as the target model. 
Equation~\ref{equ:imp} does not work against the matching-based scanner because it does not provide the confidence score to the query.
Formally, {\tech} spoofs the banner-based scanner as follows,
\begin{equation}\label{equ:rule}
\begin{aligned}
        F_{re}(\cdot) \in S_{re},  &  \quad F_\theta(x) \approx F_{re}(x)  \\
\end{aligned}
\end{equation}
where $S_{re}$ is the data set of the regex expressions, $F_{re}(\cdot)$ is a string matching module, and $F_\theta(x)$  is a shadow model to approximate $F_{re}(x)$, as the shadow model. 
Here, we directly utilize a shadow model to find a perturbation in the IoT banner.

\begin{algorithm}[!t] \small
    \SetKwData{Left}{left}
    \SetKwData{This}{this}
    \SetKwData{Up}{up}
    \SetKwFunction{Union}{Union}
    \SetKwFunction{FindCompress}{FindCompress}
    \SetKwInOut{Input}{Input}
    \SetKwInOut{Output}{Output}
    \Input{Shadow model $F$, original $x$, label $y$;  \\
           Substitution space $C$,  IoT-semantic space $S_{sem}$ }
    \Output{Adversarial example $x_{adv}$}
    \BlankLine
    $y \leftarrow F(x)$;  pick $S_y$ from $S_{sem}$;\\
    $x' \leftarrow x$; \\
    \For{ $w_i \in x$ \& $F(x') = y$ }{
        \If{$w_i \in S_y$}{
            \While{$C(w_i) \neq \emptyset $}{
                $w'$ = pick one ($C(w_i)$);   \\ 
                $C(w_i).delete(w')$; \\
                \If{ $w' \notin S_y$}{
                    $x' \leftarrow x'_{w_i\leftarrow w'}$;\\
                    break; \\
                }
            }
            
        }
    }
    $x_{adv} \leftarrow x'$; \\
\Return $x_{adv}$
\caption{ \textit{{\tech}}: Generating adversarial examples against a matching-based scanner. }
\label{alg:rule:generated}
\end{algorithm}

\textbf{Shadow Model} 
We can use the machine learning algorithm to train a shadow model.
The shadow model is not the targeted model, and we cannot use it to replace the matching-based scanner.
Instead, we leverage the shadow model to obtain a label $y$ of an IoT banner $x$.
After obtaining the label $y$, we leverage the IoT-semantic set $S_y$ to find the available perturbations for adversarial examples.

\textbf{Similarity}.
We utilize cosine similarity to find a similar word in the IoT semantic set $S_y$ (see Section~\ref{sec:sub:sem}).
We employ a heuristic rule to discover a similar word in the IoT-semantic set $S_y$: if two words exist in the same IoT banner, two words have a high correlation.
Specifically, we use the loss function to minimize the difference between the predicted similarity and the true similarity, as follows,
\begin{equation}\label{equ:loss}
    J=\sum_{i=1}^{|V|}\sum_{j=1}^{|V|}f(P_{(i,j)})(\theta_i^T\theta_j+b_i+b_j-logP_{(i,j)})^2
\end{equation}
where $|V|$ is the size of the vocabulary, $P_{(i,j)}$ is the number of co-occurrences of words $i$ and $j$, $\theta_i$ and $\theta_j$ are the vector representations of words $i$ and $j$, $b_i$ and $b_j$ are bias items, 
$f(x)$ is a weight function to adjust the impact of the number of the pair ($i$, $j$) to the loss function.

Algorithm~\ref{alg:rule:generated} depicts the overall process of generating adversarial examples of IoT device banners. 
Given a banner $x$, the shadow model finds its substituted set $S_y$ in the IoT semantic space. 
In each iteration, we only pick one substituted word whose similarity is larger than 0.8. 
Once a qualified example is generated, we use it against the matching-based scanner.

\begin{algorithm}[!t] \small
    \SetKwData{Left}{left}
    \SetKwData{This}{this}
    \SetKwData{Up}{up}
    \SetKwFunction{Union}{Union}
    \SetKwFunction{FindCompress}{FindCompress}
    \SetKwInOut{Input}{Input}
    \SetKwInOut{Output}{Output}
    \Input{ Banner set $X = \{x_1, ... x_m \}$, Label Set $Y=\{y_1, y_2, \dots, y_n \}$ \\
    }
    \Output{ IoT Semantic Set $S_{sem} = \{S_1, ... , S_n\}$ }
    \BlankLine
    $S_{sem} \leftarrow \{ \};$  \\
    \For{ $y \in Y$}{
        $S_{y} \leftarrow []$; 
    }
    \For{$x\in X$}{
        $y \leftarrow F(x)$;  \Comment{$x$ is $\{w_1, w_2, \dots w_n\}$}  \\
        $W_{l}   \leftarrow  Sort(\{w_1, w_2, \dots w_n\}$)  \textit{by Equ.~\ref{equ:imp}} \\ 
        $S_y$.append($W_l$);\\
    }
    \For{$y\in Y$}{
        Sort($S_y$, TF);  \\
        remove duplicated in $S_y$; \\
        $S_{sem}$.add($S_y$); \\
    }
    
    \Return $S_{sem}$;
\caption{ Generating IoT Semantic Space. }
\label{alg:generated:ruleset}
\end{algorithm}

\section{Perturbation Space Generation}
\label{sec:space}

Perturbation space represents the substitution set for perturbing a word in IoT device banners, including an IoT semantic space and a visual similarity space.

\subsection{IoT Semantic Space}

In the field of natural language processing (NLP), there are many existing works to find a semantic substituted set, including word embedding-based method~\cite{sato2018interpretable}, language model-based method~\cite{zhang2019generating}, synonym-based method~\cite{ren2019generating}, and the sememes dictionary~\cite{dong2010hownet}. 
However, those candidate substituted sets in the IoT banner differ from the corpus under the NLP field.
First, many IoT-related words belong to non-dictionaries and never appear in NLP dictionaries. 
Second, many IoT-related words are similar but irrelevant to each other. 
For example, ``dcs-930l'' and ``D-Link'' should have a similar semantic relationship because ``dcs-930l'' is an IoT device product of the manufacturer ``D-Link''.
If we directly apply the corpus (e.g., WordNet~\cite{sato2018interpretable, zhang2019generating} or HowNet~\cite{dong2010hownet}) to pick candidate substitutes for a perturbation, the IoT device banner has semantic sparsity, decreasing the performance of the adversarial example.

To address this issue, we build a semantic space containing various IoT-related keywords.
Specifically, We leverage the following equation~\ref{equ:imp} to calculate the confidential score of every word in the IoT banner and pick the top $l$ words for constructing the semantic substitution set.
Then, we use Term Frequency (TF) to sort those IoT-related words.
TF refers to the number of IoT-related words appearing in the dataset. 
A high frequency indicates that the word is crucial to the classification predictions. Therefore, we leverage the TF to sort those IoT-related words from high to low for each device label. 
Note that there are duplicated words in the IoT semantic substitution set, and we would remove those duplicated ones.
After that, we generate the IoT-related semantic space for the IoT device banners dataset.
Typically, we pick candidate substitutes for each word in the IoT device banner with the top closest synonyms in semantic space.

Algorithm~\ref{alg:generated:ruleset} lists the process of constructing IoT semantic space, where the input is the banner set ($X$) and label set ($Y$), and the output is the IoT semantic set ($S_{sem}$).

\begin{figure}[!t]
  \centering
  \includegraphics[width= 3.2in]{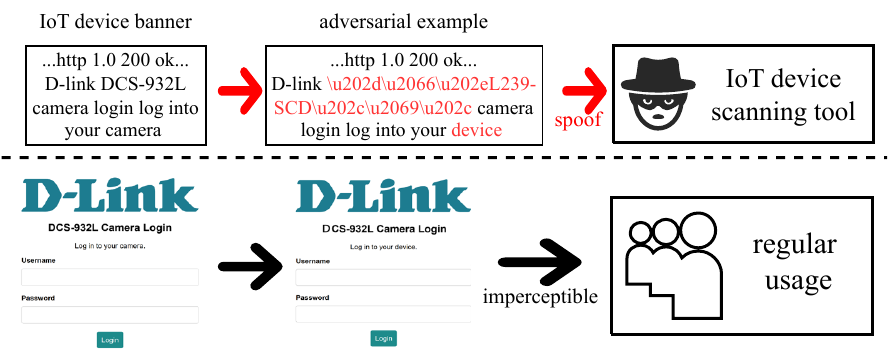}
  \caption{Visual consistency based on Unicode for adversarial IoT banners.}
  \label{fig:consideration}
\end{figure}

\subsection{Visual Similarity Space}
\label{sec:sub:unicode}

When processing IoT banners, there is a difference between humans' eyes and programs: the program only considers the encoding and semantics of the input without the input's visual appearance.
For humans, the visual rendering effect plays a decisive role in human understanding of the meaning of the text. 
Visual modifications are sufficiently subtle to go unnoticed by average users.
The Unicode specification~\cite{unicode} is a universal character encoding standard used to assign unique digital codes to computer characters.
Unicode includes a variety of encoding methods. 
The most commonly used is UTF-8, which has better compatibility and the characteristics of variable length encoding.
Figure~\ref{fig:consideration} depicts the Unicode perturbations in an IoT device banner that deceive the IoT device scanning tool but are imperceptible to humans without affecting regular usage.

Based on the Unicode character set, we use three operations to represent the visual similarity space. 
(1) Visual similarity character. 
We propose two character sets, the Greek alphabet ($G$) and the Russian alphabet ($R$), as the rendering set. The visual operation is to replace the original characters with the corresponding characters in the rendering set.
For example, the Latin character ``a'' (U+0061) and the Russian character ``a'' (U+0430) are visually similar, where users may not notice their differences.
(2) Zero-width space character (U+200B). 
We propose to insert a zero-width space character as the visual operation that inserts a word's position for generating word segmentation.
This character is invisible to humans but can split one word into two. After doing this, the IoT scanner would convert them into two tokens and two embedding vectors.
(3) Reversed order character. 
The program rigorously follows the character encoding order when reading input. Any change in the encoding order of characters transforms that word into a distinct one.
Given a word in the IoT banner, we reverse it by inserting the character (U+202C) and then adding the character (U + 202E), displaying it correctly visually. 
A detailed unicode reversed character set is listed in Table~\ref{tab:unicode} in the Appendix.

Take an example. 
Figure~\ref{fig:unicode:html} in the Appendix depicts an HTTP banner from an IoT device between before and after the visual perturbation. 
Figure~\ref{fig:unicode} in the Appendix depicts an FTP banner from an IoT device between before and after visual perturbation.

\section{Evaluation}
\label{sec:eva}

In this section, we first present the implementation detail of {\tech} and experimental settings. 
Then, we evaluate the performance of {\tech} and compare it with the baseline approaches.
Further, we shed light on analyzing the scanners' weaknesses in the carefully crafted sample.

\begin{figure}[!t]
	\centering
	\includegraphics[width= 3.2in]{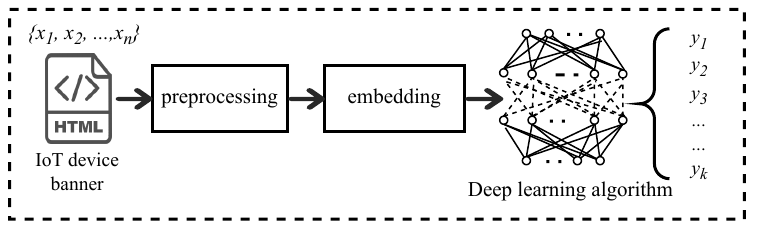}
	\caption{A learning-based IoT scanner.}
	\label{fig:back:model}
\end{figure}

\begin{figure}[!t]
  \centering
  \includegraphics[width= 3in]{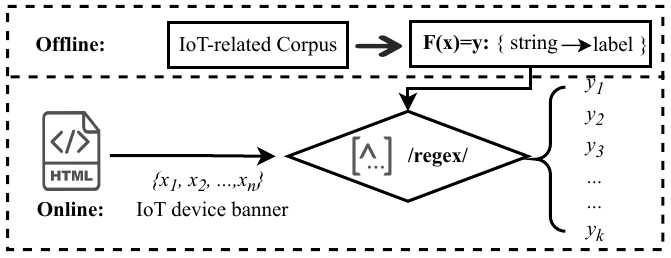}
  \caption{A matching-based IoT scanner.}
  \label{fig:back:rule}
\end{figure}

\subsection{Implementation}

\textbf{A Learning-based Scanner} infers the function $F(\cdot)$ between the banner $x$ and the corresponding device label $y$.
Figure~\ref{fig:back:model} depicts the overall of the learning-based scanner.
The banner $x$ is the packet payload in the application-layer protocol.
We need to remove redundant and irrelevant elements in the banner $x$ and obtain a clean banner that contains the essential information necessary for the learning-based scanner; 
Then, We can convert the token into the corresponding word embedding vector through the embedding technique.
We utilize the deep neural network architecture to infer the classification model for IoT devices, where the input is the vector and the output is the device label.

We have utilized the NLTK tool~\cite{nlp} to segment and extract the word from the banner for word tokenizations. We have used the PyTorch library~\cite{pytorch} to implement neural network algorithms, including the Convolutional neural network (CNN) and the Long short-term memory (LSTM) network. 
The learning-based scanner acts as a classification task, where its model is a multi-classifier. 
Table~\ref{tab:model:parameter} (Appendix) lists the parameters of selected target models for detecting IoT devices.

\textbf{A Matching-based Scanner} is to leverage string matching to map the banner $x$ into the device label $y$.
Figure~\ref{fig:back:rule} depicts the overall of the matching-based scanner for IoT devices.
In the offline stage, we construct an IoT-related corpus to store useful keywords in the database. 
In the online stage, we use the regex expression to match the IoT banner to find and collect IoT device information. 

We have extracted existing regex expressions from prior works, including Nmap~\cite{nmap}, Zmap~\cite{durumeric2013zmap}, and ARE~\cite{xuan2018are}.
We use a unified form of \{regex/string$\rightarrow$label\} to present the matching-based scanner, where the regex/string is the IoT-related keyword, and the label is the IoT device information. 
Table~\ref{tab:rule:num} lists the number of keywords of the matching-based scanner for deriving IoT device labels. Note that these scanners employ distinct formats and device label name schemes, we provide manual calibration to those regex expressions and strings.
For the matching-based scanner, we store all regex/strings in the JSON file and utilize string matching to identify IoT devices. 
We have used the Yake library~\cite{yake} to implement the keyword extraction and regex matcher.

\textbf{{\tech} Implementation}.
We have implemented a prototype system as a self-contained piece of software based on open-source libraries, including PyTorch framework~\cite{pytorch} and TextAttack~\cite{morris2020textattack}.
We have used the TextAttack library to implement two evasion strategies.
We have written a custom Python script to connect two perturbation spaces and evasion strategies for generating adversarial examples of IoT banners.
For the IoT-semantic space, the top $l$ words in each category is 20. 
For the visual similarity space, we use the regex for HTTP division in Table~\ref{tab:region} (Appendix). 
For the attack strategy against the learning-based scanner, the maximum iteration number $T$ is 20, the minimum threshold of the Jaccard similarity coefficient is 0.85, the top $K$ word is 20, the variable $a$ and $b$ are set to 0.8 and 0.2 respectively.
For the attack strategy against the matching-based scanner, the cosine similarity is equal to 0.8.
We deploy the prototype system {\tech} on a Ubuntu 18.04.6 LTS server powered by two Intel Core i9-9820X CPUs, 64GB RAM, and two 2080 NVIDIA GPUs.

\begin{table}[!t]
	\renewcommand{\arraystretch}{1.3}
	\caption{The number of IoT device rules and corresponding performance.}
	\centering
	\begin{tabular}{ c c  }
	\toprule
		Tool  &  Number of IoT-related keywords  \\
	\midrule
		Nmap~\cite{nmap}/Zmap~\cite{durumeric2013zmap}	&  287 regex    \\
		ARE~\cite{xuan2018are} 	&  37,903 strings   \\
	\bottomrule
	\end{tabular}
	\label{tab:rule:num}
\end{table}

\subsection{Experimental Settings}

\textbf{Dataset}.
We have collected IoT device banners through the following process.  
Those IoT device banners are collected through the network scanning activity, where we send the probe to port 80/23 of remote hosts in cyberspace and build 3 handshakes for collecting HTTP and FTP application packet payloads.
For each banner, we use three open-source tools (Nmap~\cite{nmap}, Zmap~\cite{durumeric2013zmap}, and ARE~\cite{xuan2018are}) to obtain device labels.
If a banner does not return any IoT-related information, we only remove it.
We leverage the Shodan~\cite{Shodan} to conduct cross-validation on labels of those banners.
If their device labels are the same, we store them in the dataset of IoT device banners; otherwise, we drop them. 
We have collected 179,000 IoT device banners as the dataset.  
Table~\ref{tab:dataset} lists the dataset for IoT devices, including 10 device types, 41 device manufacturers, and 692 device products.
We have covered popular device types, such as routers, switches, cameras, and printers. 
Our dataset is imbalanced for device manufacturers and products, where there is a long tail of 179,000 banners, while the top 10 manufacturers produce 70\% of the device banners.

\begin{table}[!t]
	\renewcommand{\arraystretch}{1.3}
	\caption{The dataset for IoT device banners}
	\centering
	\begin{tabular}{ c c }
	\toprule
		Category &  Number of Device Class \\
	\midrule
		Device Type 	&  10 \\
		Device Manufacturer 	&  41 \\
		Device Product 	&  692 \\
	\bottomrule
	\end{tabular}
	\label{tab:dataset}
\end{table}

\textbf{Metric}. 
We use the following metrics to measure the effectiveness of {\tech}.
\begin{enumerate}
    \item Success Rate (SR) measures the proportion of adversarial examples that spoof the IoT device scanner. 
    In contrast to the accuracy rate, the high SR refers to the effectiveness of the adversarial examples and incorrect discrimination on the IoT banners.
    \item Modified Range (MR) indicates the different degree between the adversarial example and the original one.  
    Here, we use the Jaccard distance to measure the MR value.
    Large MR indicates that the adversarial example is far from the original banner. 
    \item Visual Consistency (VC).
    After the visual rendering system, adversarial examples should be consistent with the original IoT banner from the user's perspective. 
    \item Query Number (QN). 
    In the black-box setting, the QN measures the cost of generating adversarial examples. 
    One query indicates that we can obtain the targeted model's confidence score for a perturbation in the IoT banner.   
\end{enumerate}

\subsection{Performance against Learning-based Scanner}

We first validate the performance of the learning-based scanner for recognizing IoT devices, where the input is the banner and the output is the device label. 
Table~\ref{tab:perf:model} lists the performance of the CNN and LSTM models, including device type level, manufacturer level, and product level. 
On average, the two classification models perform similarly for the learning-based scanner. 
For the device type, the CNN has 98.32\% accuracy, and LSTM  has 98.15\%.
For the device manufacturer, the CNN has 97.17\% accuracy, and LSTM has 97.08\%.
Comparatively, the CNN achieves an accuracy of 95.50\%, while the LSTM reaches 94.53\% at the device product.
The reason is that the multi-classifier's performance is affected by the number of categories, where the number of product labels is much larger than the number of device types and manufacturers.
In short, we use their performance as a starting point to generate adversarial examples of IoT banners.

\begin{table}[!t]
	\renewcommand{\arraystretch}{1.3}
	\caption{The performance for the learning-based scanner.}
	\centering
	\begin{tabular}{ c c c c c}
	\toprule
		           &  Device  Type   & Device Manufacturer & Product \\
	\midrule
		CNN     	&  98.32\%  & 97.17\% & 95.50\%  \\
		LSTM      &  98.15\%  &  97.08\%   &   94.53\% \\ 

	\bottomrule
	\end{tabular}
	\label{tab:perf:model}
\end{table}

\begin{table}[!t]
	\renewcommand{\arraystretch}{1.3}
	\caption{The performance for {\tech} against the learning-based scanner. }

	\centering
	\begin{tabular}{ c c c c c }
	\toprule
		            &  SR        		 &    MR  			& QN  		 & VC \\
	\midrule
	Device Type 	&  66.89\%  			& 0.134		& 158.72   &    Yes \\
	Device Manufacturer 	&  77.56\%  		& 0.144 		& 193.57   &   Yes \\
	Device Product		&  80.84\%  		& 0.066 		&100.66    &    Yes \\

	\bottomrule
	\end{tabular}
	\label{tab:model:Odsay}
\end{table}

We conducted the validation experiments of {\tech} against the learning-based scanner. 
Table~\ref{tab:model:Odsay} lists the performance of the adversarial examples, including device type, manufacturer, and device product.
Results indicate that the success rate (SR) for spoofing IoT banners achieves 66.89\% for device type, 77.56\% for manufacturer level, and 80.84\% for device product level. 
For the modified range (MR), we find that the MR is 0.134 at the device type level, 0.066 at the product level, and 0.144 at the manufacturer level. 
For the cost, we observe that the query number is the largest in the device manufacturer level, close to 193.57, followed by the device type level 158.72 and the product level 100.66. 
For visual consistency (VC), we utilize manual inspection to determine whether an adversarial example changes the appearance of its original banner. 
We cannot inspect all banners to determine whether they are visually consistent.
Hence, we have randomly picked up several banners from each IoT device category. Finally, we manually reviewed 1,000 adversarial banners throughout this validation stage and found that those adversarial examples of banners attained visual consistency.

We provide the analysis of adversarial examples for the learning-based scanner.
First, adversarial examples are hardest generated at the device type level yet have the lowest SR, high MR, and high QN.
In contrast, adversarial examples are easily generated at the product level, with the highest SR, the lowest QN, and the minimum MR. 
The performance against the learning-based scanner at the manufacturer level is somewhere in between them.
The probable reasons are two aspects: (1) the category number is related to the robustness of the learning-based scanner; (2) the intricate features and complex patterns in the learning-based scanner are vulnerable to adversarial manipulations of the inputs. The easier it is to generate adversarial examples, the more brittle the learning-based scanner in IoT device detection. 
Overall, the granularity of the learning-based scanner is inversely proportional to the robustness degree: Type-level $>$ Manufacturer-level $>$ Product-level.

\begin{table}[!t]
	\renewcommand{\arraystretch}{1.3}
	\caption{The comparison between our {\tech} and other 2 approaches, including probability heuristic substitution and TEXTFOOLER. }
	\centering
	\begin{tabular}{ c c c c c }
	\toprule
		           &  SR         & MR  & QN    &  VC \\
	\midrule
	{\makecell{Probability \\ Heuristic~\cite{alzantot2018generating} }}	&  52.02\%  		&  0.103 		& 34.15   &    No \\
	TEXTFOOLER~\cite{jin2020bert}		&  60.89\%  	&  0.091 		& 23.42   &   No \\
	{\tech}		&  72.83\%  		& 0.055 		& 131.78   &   Yes \\
	\bottomrule
	\end{tabular}
	\label{tab:basline:l}
\end{table}

\textbf{Comparison}.
So far, there is no prior work to generate adversarial samples for spoofing IoT device banners.  
Hence, we have implemented 2 approaches in the NLP tasks as the baselines to compare with {\tech}. 

(1) Probability heuristic method~\cite{alzantot2018generating}. 
We use the importance score to calculate the position of words in the banner. 
During each iteration, every position holds a probability value, and a word is chosen to be perturbed based on its respective probability. 
The sample with the most substantial influence on the model's confidence is preserved for the subsequent iteration.
(2) TEXTFOOLER~\cite{jin2020bert}. 
We first sort all words in the banner according to their importance scores.
The perturbation includes multiple operations (adding, deleting, and replacing) in each iteration.

\textbf{Fine-tuned baselines}.
Note that those baselines must be migrated and adapted to craft perturbations in device banners.
If we directly use them to generate adversarial examples, their accuracy is close to 0. 
Hence, we carefully fine-tuned them to achieve promising results.
First, we do not use the language pre-trained model (e.g., BERT) and semantic dictionaries (e.g., WordNet and HowNet) to find word replacement candidates for those baselines. 
They are highly related to NLP tasks - those designed for the NLP domain hardly work well on the IoT device scanning tasks. 
For a fair comparison, we employ the IoT-specific space as the dictionary for the baseline methods. When substituting words, we utilize word embedding similarity to identify a suitable replacement from the IoT-specific space.
Second, we use similar parameters as {\tech}: the maximum number of iterations is set to 20, and the minimum Jaccard similarity coefficient is set to 0.85.

Table~\ref{tab:basline:l} lists the comparison results between our {\tech} and the other two approaches. 
{\tech} achieves an SR of 75.05\% and an MR of 0.055, which is superior to other methods. 
For the cost, {\tech} has the highest QN, nearly 135.45, compared with baselines. 
In the same number of iterations, the probability heuristic approach has the lowest SR and the largest MR, and TEXTFOOLER has the lowest QN. 
The probability heuristic approach yields a 0.103 MR, whereas TEXTFOOLER demonstrates a 0.091 MR. This discrepancy can be attributed to the fixed direction of perturbation. Each perturbation is stacked upon the previous one without adapting to varying directions, leading to an enhanced MR for adversarial examples.
Further, the adversarial samples generated by {\tech} go unnoticed by the human eye. 
In contrast, baselines only use the IoT-semantic to generate disturbances in device banners without guaranteeing their visual consistency.
For example, two products from the same manufacturer have similar semantic information, ``dcs-930l'' and ``dcs-932l'', and replacing those two words leads to visual inconsistency.
Results demonstrate that IoT device scanners are vulnerable to adversarial manipulations of their banners under the IoT-related dictionary.

\subsection{Performance against Matching-based Scanner}

\begin{table}[!t]
	\renewcommand{\arraystretch}{1.3}
	\caption{The performance for {\tech} against the matching-based scanner. }
	\centering
	\begin{tabular}{ c c c c }
	\toprule
		           &  SR         & MR    				& CV \\
	\midrule
	Device Type	 	&  64.60\%  		& 0.302		   &    Yes \\
	Manufacturer 	&  68.85\%  		& 0.225 		   &    Yes \\
	Device Product	&  77.71\%  		& 0.103	  &    Yes \\
	\bottomrule
	\end{tabular}
	\label{tab:rule:Odsay}
\end{table}

As mentioned, we cannot send the query to the matching-based scanner to obtain the importance score of adversarial examples. 
Hence, we do not use the query number (QN) to present the performance of {\tech}. 
Table~\ref{tab:rule:Odsay} lists the overall performance of adversarial examples against the matching-based scanner over device type, manufacturer, and device product. For the SR, it is evident that  {\tech} is the most effective in interfering with the matching-based scanner at the product level, followed by the device manufacturer and device type, as 77.71\%, 68.85\%, and 64.60\%, respectively. 
We can observe that disturbing the matching-based scanner has a lower average SR than disturbing the learning-based scanner (combining Table~\ref{tab:model:Odsay} and Table~\ref{tab:rule:Odsay}).
There are two possible reasons. First, the shadow model has some deviation from the match-based scanner.
Second, we have used many regex/strings from existing works, which seem more robust than the learning-based scanner.

{\tech} has the most modified range at the device type level (0.302), followed by the manufacturer and product level (0.225 and 0.103).  
The scanner at the device product level is the most robust, and the scanner at the type level are most frangible for adversarial examples. 
We need more cost to modify the original banner in the device type than in others. 
For the CV, our {\tech} guarantees the consistent appearance of the original banner after rendering.

\textbf{Comparison}.
As so far, existing approaches (e.g., TEXTFOOLER~\cite{{jin2020bert}}) cannot work against the matching-based scanner because there is no targeted model where we can send a query to obtain the importance score of the adversarial example. 
The matching-based scanner is a complete black box for defenders.
Hence, we use 3 baselines to compare with {\tech}.
(1) Random Changing. 
We randomly select a place in the device banner and, using the embedding similarity, pick the most similar word in the IoT semantic space.
(2) Regex Changing.
We use the regex from the Nmap/Zmap/ARE tools to select all possible places in the device banner. 
In each iteration, we pick the most similar word in the IoT semantic space through the embedding similarity.
(3) {\tech} against the learning-based scanner. 
We directly use the model for the adversarial samples and migrate them to spoof the matching-based scanner.

Table~\ref{tab:com:model} compares {\tech} and 3 baselines.
It is obvious that the ${\tech}_{matching}$ has achieved the best results, the highest SR. 
${\tech}_{learning}$ has the lowest SR and MR.
We claim that adversarial examples generated by the learning-based scanner cannot be migrated to the ones against the matching-based scanner.
Random Changing makes many useless perturbations in the banners, leading to the largest 0.091 of MR and 52.02\% SR.
Regex changing makes a high SR and a high MR.
The reason is that an IoT banner has many places matched with the regex, leading to a larger MR value. 
In addition, the ${\tech}_{matching}$ and ${\tech}_{learning}$ both guarantee the visual consistency for IoT banners.

\begin{table}[!t]
	\renewcommand{\arraystretch}{1.3}
	\caption{The comparison between our {\tech} and baselines, including random changing, regex changing, and ${\tech}_{learning}$. }
	\centering
	\begin{tabular}{ c c c c c }
	\toprule
		           &  SR         & MR    & VC \\
	\midrule
	Random Changing     	&  52.47\%  		& 0.224 		  &    No \\
	Regex Changing    &  76.29\%  	&0.109   		   &   No \\
	{\tech}$_{learning}$		&  16.56\%  	&  0.055 		   &   Yes \\
	{\tech}$_{matching}$		&  77.71\%  		& 0.102 	   &   Yes \\
	\bottomrule
	\end{tabular}
	\label{tab:com:model}
\end{table}

\subsection{Analysis of IoT Scanner's Weakness}

\begin{table}[!t]
	\renewcommand{\arraystretch}{1.3}
	\caption{The analysis between semantic changing and visual rendering changing for disturbing IoT fingerprints. }
	\centering
	\begin{tabular}{ l c c c  }
	\toprule
		           &  MR         & QN     & VC \\
	\midrule
	{\tech}$_{learning}$ + Visual       	&  0.067 		&  173.49	  &    Yes \\
	{\tech}$_{learning}$ + Semantic	        &  0.055   	&  132.88		   &    No \\
	{\tech}$_{learning}$ + both		        &  0.056   	&  135.45		   &   Yes \\
	\bottomrule
	\end{tabular}
	\label{tab:ana:model}
\end{table}

First, we analyze the IoT semantic and visual similarity spaces to find perturbations in IoT banners.
Table~\ref{tab:ana:model} lists the comparison performance for {\tech}$_{learning}$, including MR, QN, and VC. 
The semantic similarity perturbation is more efficient than the visual rendering perturbation, with a 0.055 modified rate and a 132.88 query number. 
The reason is that the semantic space can leverage the context information of the IoT banner for finding a perturbation, but the visual similarity space does not have the banner context.
The visual similarity space has the advantage of maintaining visual consistency but contains a larger MR and QN.
It is clear that the {\tech}$_{learning}$ achieves the best performance when we use both the IoT semantic space and visual similarity space.

\begin{figure}[!t]
    \centering
    \includegraphics[width=2.2 in]{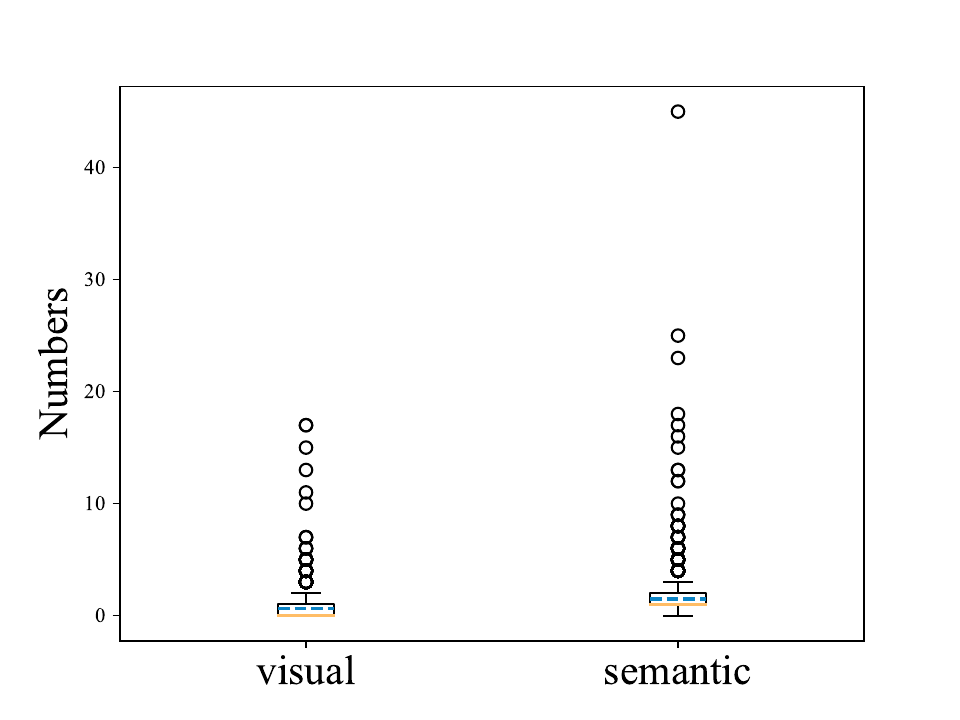}
    \caption{Distribution of visual and semantic perturbation number in adversarial examples of IoT banners.}
    \label{fig:in-sem}
\end{figure}

We analyze the perturbation number in adversarial examples of IoT banners, as shown in 
Figure~\ref{fig:in-sem}.
In the visual space, the average perturbations in adversarial examples are 0.6; in the semantic space, adversarial examples have more than 1.48 perturbations on average. 
The perturbation number in adversarial examples has a long tail, indicating that those IoT banners require dozens of perturbations to spoof IoT scanners. 
We analyze the IoT-semantic space for each device category.
Table~\ref{tab:ana:rule} lists the detailed IoT-semantic space.
We have 692 device categories of IoT devices, where each category has nearly 18.48 related keywords on average. 
The size of the IoT-semantic space is 692KB.

\begin{table}[!t]
	\renewcommand{\arraystretch}{1.3}
	\caption{ The IoT-semantic space for IoT device banners. }
	\centering
	\begin{tabular}{ c c  }
	\toprule
		           &  Number          \\
	\midrule
	Category Number        	&  692  		 \\
	Words per Category	        &  18.48 avg.   	 \\
	Size 		        &  692kb   	\\
	\bottomrule
	\end{tabular}
	\label{tab:ana:rule}
\end{table}

\begin{figure*}[!t]
    \begin{tabular}{ c   c  c }
    \begin{minipage}[t]{0.31\linewidth}
    \centering
    \includegraphics[width=1.0\linewidth]{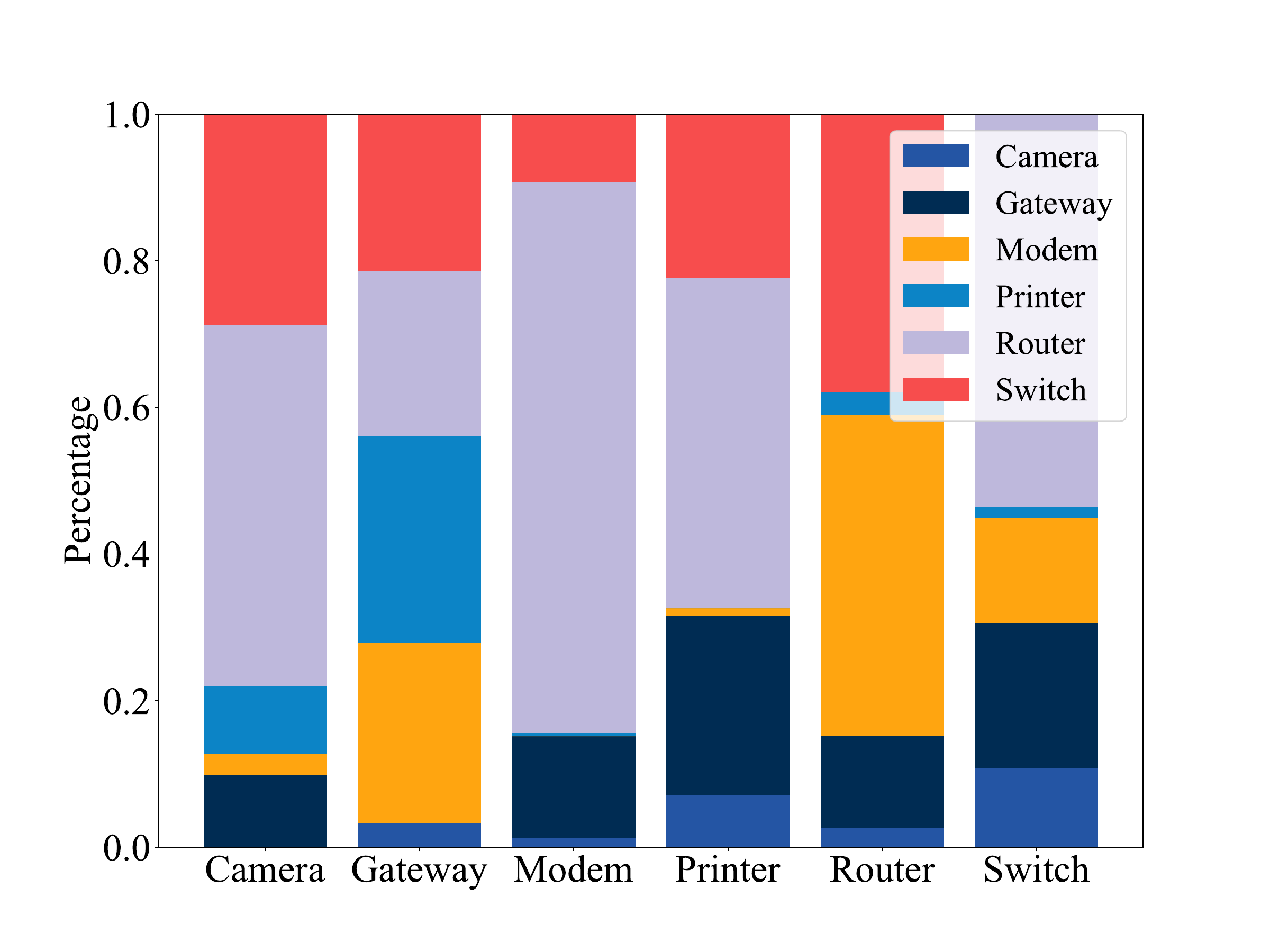}
    \caption{Device-type level: erroneous predictions of adversarial examples.}
    \label{fig:dt:sr}
    \end{minipage}
    &
    \begin{minipage}[t]{0.31\linewidth}
     \includegraphics[width=1.0\linewidth]{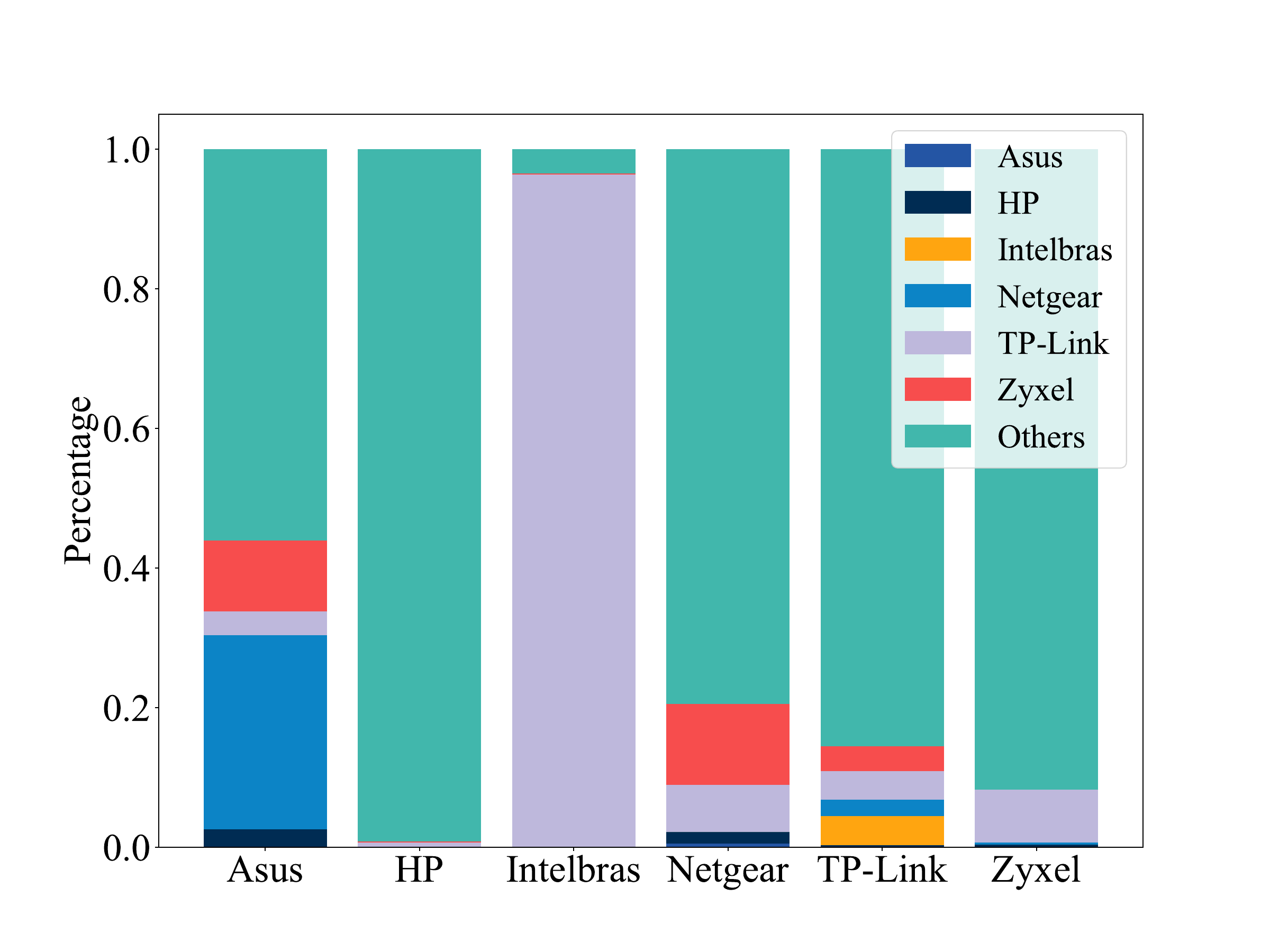}
    \caption{Manufacturer level: erroneous predictions of adversarial examples.}
    \label{fig:manufacturer:sr}
    \end{minipage}
	&
    \begin{minipage}[t]{0.31\linewidth}
    \includegraphics[width=1.0\linewidth]{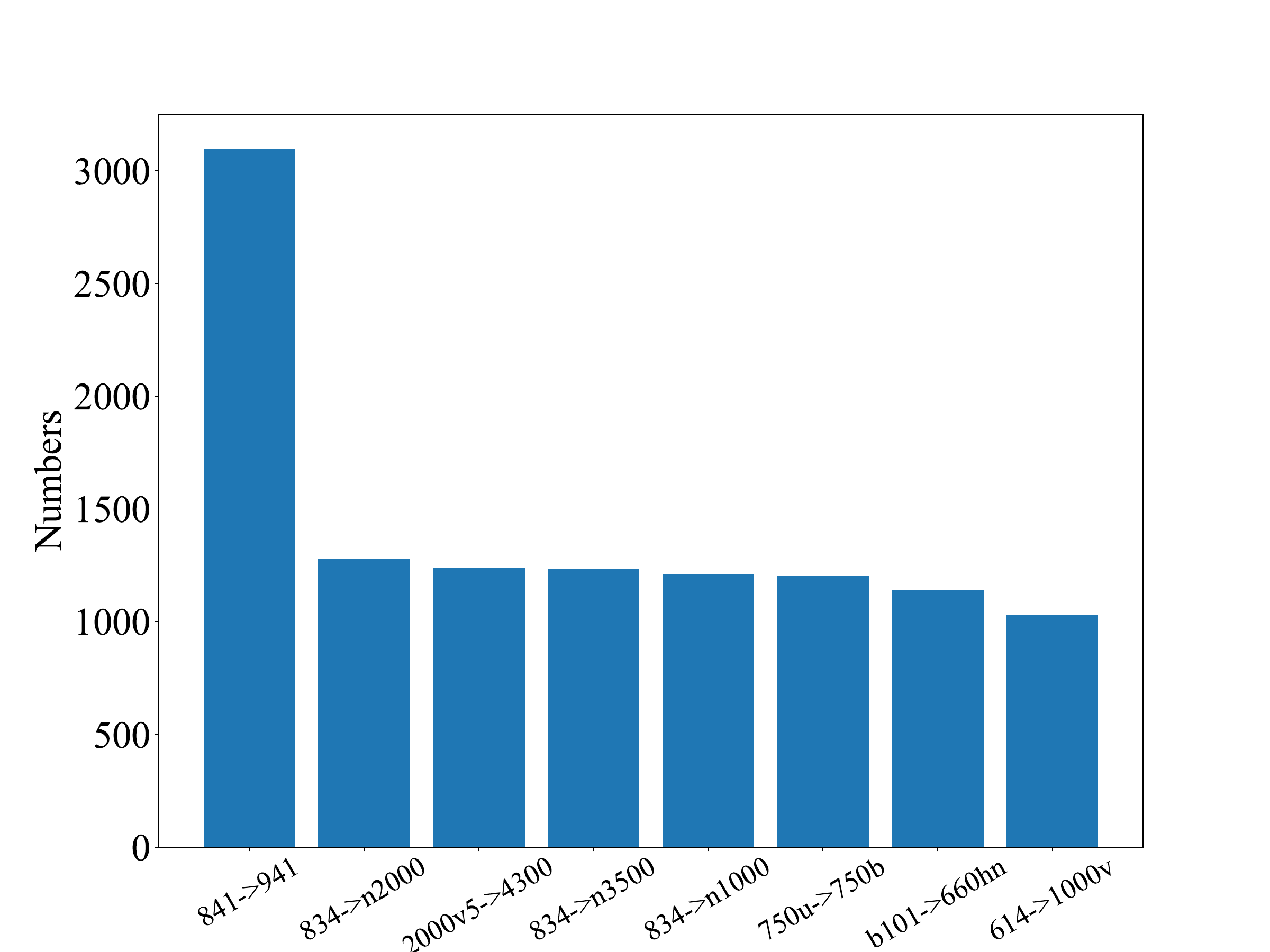}
    \caption{Product level: erroneous predictions of adversarial examples.}
    \label{fig:product:sr}
    \end{minipage}\\
    \end{tabular}
\end{figure*}

\begin{figure}[hbt]
    \centering
    \includegraphics[width=2.4in]{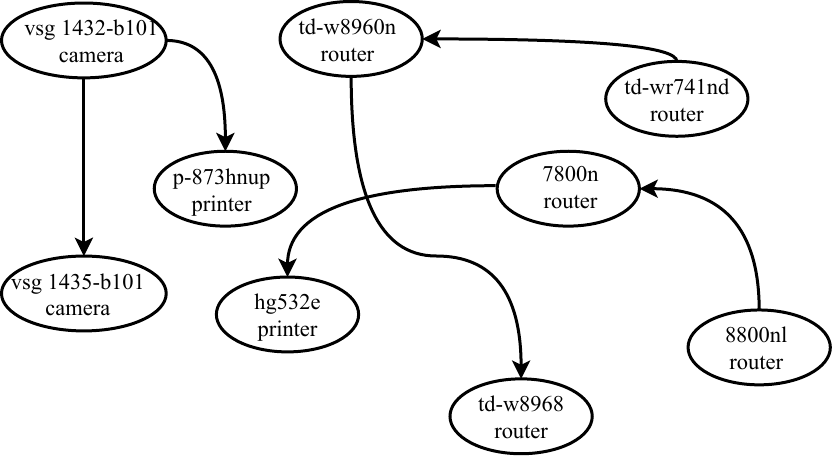}
    \caption{ $A\rightarrow B$: $A$ is the original label of the IoT banner, and $B$ is the erroneous prediction of its adversarial example.}
    \label{fig:state:mis}
\end{figure} 

Further, we analyze erroneous predictions of adversarial IoT banners classified by IoT scanners.
In the device type level, Figure~\ref{fig:dt:sr} depicts the distribution of adversarial banners misclassified by IoT scanners.
The X-axis presents the original device type, and the Y-axis presents the percentage of erroneous predictions.
For IoT scanners, other device types (e.g., cameras, modems, printers, and switches) can easily be spoofed as router types. 
Additionally, the router type of an IoT banner is likely to be misclassified as the modem or printer type. 
We find that the print and camera types are the least predictions of adversarial IoT banners from other types.
The reason is that their banners differ greatly among those device types, leading to few adversarial banners. 

Figure~\ref{fig:manufacturer:sr} shows the erroneous predictions of adversarial IoT banners at the manufacturer level, where the X-axis presents the original manufacturer label, and we pick 6 popular manufacturers (Asus, HP, Intelbras, Netgear, TP-Link, Zyxel). Note that we have 41 manufacturers and use the label ``other'' to denote the remaining manufacturers.
Results demonstrate that 90\% of devices from the manufacturer ``Intelbras'' are misclassified into the manufacturer ``TP-Link'', and nearly 100\% of devices from the manufacturer ``HP'' are misclassified into ``other''.
More than 50\% of adversarial banners are misclassified into the ``other'' category.

Figure~\ref{fig:product:sr} shows the erroneous predictions of adversarial IoT banners at the product level, where we pick the top 8 pairs (original$\rightarrow$changed).
We use an example to illustrate the pair (original$\rightarrow$changed) as shown in Figure~\ref{fig:state:mis} (Appendix).
Note that there are 692 device product labels and 478,172 misclassified pairs (692*691).
We find that many adversarial IoT banners are misclassified within the same device type and manufacturer. 
The product ``tl-wr841nd'' comes from the manufacturer ``TP-Link'', and its adversarial banner is misclassified into the product ``tl-wr941nd'' from the same manufacturer.
The pair (tl-wr841nd$\rightarrow$tl-wr941nd) contains the largest number of IoT banners, nearly 3,096.
In contrast, we find that some adversarial IoT banners are misclassified into different device types and manufacturers.
The product ``vsg1432-b101'' belongs to the camera type, but its adversarial banner is misclassified as the product ``p-660hn'', which belongs to the printer type.
Those adversarial banners make the prediction results irrelevant to the original category.

\section{Discussion \& Limitations}

\subsection{Compared with Other approaches}

\begin{figure}[!t]
  \centering
  \includegraphics[width= 2.9 in]{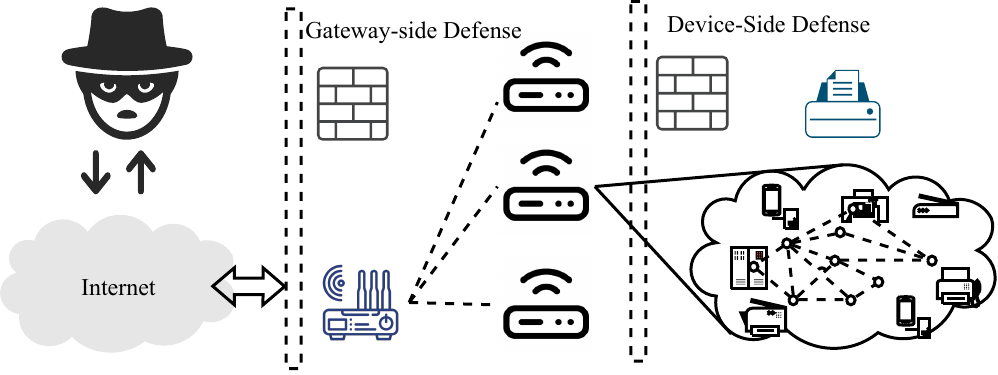}
  \caption{Defense Model: Approaches against IoT device scanning activity.}
  \label{fig:back:defense}
\end{figure}
 
So far, little prior work has been done to obfuscate IoT device scanning, and we survey relevant literature to investigate several practices from the industry and the network reconnaissance domain.
Figure~\ref{fig:back:defense} classifies the defense approaches into two categories: device-side and gateway-side.

\begin{table*}[!t]
	\caption{Comparison with State-of-the-art Studies.}
	\centering
 
        \begin{tabular}{c c c c c c c c}
		\toprule
		                      &Location  &Change Information &  \makecell[c]{Deceptive \\Information} & \makecell[c]{Influence on \\ Scanning}   &  Blocking &User perception\\
		\toprule 
		Port Close~\cite{port} &Device-side     &Port                  &No  &Yes   & Yes  &Yes \\
		MTD~\cite{sengupta2020survey, sajid2021soda}        &Device-side     &Attack Surface        &Yes &Yes  & No   &Yes \\
        CMD~\cite{mimic}        &Device-side    &Attack Surface        &Yes &Yes  & No   &Yes \\
        NAT~\cite{salutari2018closer, acar2018web}        &Gateway-side     &IP                    &NO  &Yes  & Yes  &No  \\
        Firewall~\cite{firewall,firewall-list}   &Gateway-side     &No Change             &No  &No   & No   &No  \\
        Honey-X~\cite{provos2003honeyd, honeynet-project,honey-x}    &Gateway-side   &Device Information    &Yes &No   & No   &No  \\ \hline
        {\tech}      &Device/Gateway-side     &Banner        &Yes &No   & No   &NO  \\
		\toprule
	   \end{tabular}

    \label{tab:comp}
\end{table*}

\textbf{Device-side Approaches}.
Closing port is a straightforward solution to block the application banner collection and IoT device scanning of an attacker. However,
closing port also prevents the remote access and management of IoT devices, and users cannot easily organize, configure, and maintain their devices as usual.
As an alternative, port randomization is an obfuscation technique that 
forces an attacker to guess the ports in IoT devices.
RFC 6056~\cite{ref6056} proposes to use ephemeral ports, instead of well-known ports, to mitigate malicious scanning.
However, port randomization is rarely adopted by IoT devices, and direct detection of IoT device achieves 16 million with well-known ports (see Section 6). 
Device vendors (e.g., Cisco) prevent the randomization of well-known ports that are assigned to common services in IoT devices.

Another device-side defense is to deploy moving target defense (MTD)~\cite{sengupta2020survey, sajid2021soda} on  IoT devices by dynamically changing devices' attack surface.
MTD is a proactive defense mechanism, disrupting the network reconnaissance and increasing the attack complexity.
Yet, MTD has a high cost and unsuitable for deployment on IoT devices with limited resources and computing capabilities.

\textbf{Gateway-Side Approaches}.
Network address translation (NAT) can prevent IoT scanning by mapping the internal local area network with the external wide area network. 
IoT devices behind a NAT box are not routable and invisible from the public Internet.
However, prior works~\cite{salutari2018closer, acar2018web} demonstrate that an outsider can leverage application-layer protocols or web-based vulnerabilities to find IoT devices behind the NAT.
Another observation is that popular platforms such as Censys~\cite{durumeric2015search} and Shodan~\cite{Shodan} depict millions of IoT devices accessible on the Internet.

A firewall service deployed in gateways can detect active network reconnaissance against an active port in real time.
Windows Defender Firewall~\cite{firewall} provides rules to identify network traffic and find port scans in a short time span.
In practice, such a service is often found to be in a disabled state because of perceived complexity or operational overhead.
We further found that many gateways~\cite{firewall-list} use Linux-based and monolithic firmware due to the simplicity of the Linux-stripping gateway.
HoneyNet is to deploy a series of honeypots to mimic IoT devices for deceiving the device scanning of  
The various types of software or hardware-based honeypots can be integrated into the gateway-side defense.
The false information of IoT device profiles provided by HoneyNet acts as a proactive defense.

We compare {\tech} with other other approaches, as listed by Table~\ref{tab:comp}.
Specifically, we use six general metrics to compare {\tech} and existing defense tools. 
Given the variety of existing defenses, it is practically impossible to develop a technology that can outperform them all.
Our {\tech} has several advantages: (1) being deployed on the device/gateway side; (2) only changing application banners; (3) using deceptive information to mislead attackers; (4) no affecting regular network scanning; (5) no blocking network traffic; and (6) no affecting user perception. 
In short,  we can easily deploy adversarial banners on both the device and gateway sides.
On the device side, a shell program with IPtable library~\cite{iptable} can alter IoT device banners.
On the gateway side, NAT (e.g., netfilter) uses the packet filter ruleset to revise the application packets.
Further, an adversarial banner is a lightweight approach compared with other defense mechanisms, including MTD and the honeynet.

\subsection{Limitations}

We further present {\tech}'s limitations, including the evasion and ethical considerations.

\emph{Evading {\tech}}.
Security is the battle field with endless offensive and defensive maneuvers, where two parties are always in the game.
Attackers may adopt the operations from the adversarial examples to evade {\tech}.
We further experimented to validate whether {\tech} is affected by the filtering operations by an attacker.
We assume that attackers know there are some obfuscating operations in the IoT device banners, and they can use the preprocessing module to filter out our obfuscations, e.g., removing Unicode and removing irrelevant elements and interference parts. 
We use the same dataset to evaluate {\tech}'s performance.
Table~\ref{tab:case:before} lists the accuracy of the IoT device scanner for the adversarial examples before/after the attacker carries out corresponding operations. With the help of the preprocessing operation, the accuracy of the IoT device scanner still falls quickly due to the use of {\tech}. 
So far, we have not considered how to migrate the adversarial training when attackers can utilize adversarial examples in their scanner tools.

\begin{table}[!t]
	\renewcommand{\arraystretch}{1.3}
	\caption{{\tech}'s performance before/after attackers remove our obfuscation operations.}
	\centering
	\begin{tabular}{ccccccc}
	\toprule
		{\bf Num. of Pert.}	& {\bf 2}& {\bf 4}&{\bf 6}& {\bf all}\cr
		
		\midrule

		{\bf Before Acc.}		&37.89$\%$&27.21$\%$&25.15$\%$&22.29$\%$\cr
		{\bf After Acc. }&50.05$\%$&47.38$\%$&46.72$\%$&46.11$\%$\cr
		
	\bottomrule
	\end{tabular}
	\label{tab:case:before}
	\end{table}

\emph{Ethical Considerations}.
While {\tech} can act as a tool for maintaining privacy and thwarting malicious scanning activities, ethical considerations need to be taken into account when applying the obfuscation technique.
First, {\tech} is also employed to deceive users or authorities, such as evading security measures or concealing illegal activities. 
Second, when applying {\tech}, it is important to notify all users that their devices are being obfuscated and provide the reasons behind them.
Third, adversarial examples may pose challenges to security researchers who rely on scanners and fingerprints to identify underlying threats.
In general, achieving specific goals while maintaining ethical principles is essential when applying {\tech} for obfuscating IoT device scanners.


\section{Related Work}
\label{sec:related}

\textbf{IoT device measurement} is to leverage network scanning to collect IoT devices on the Internet. 
Kreibich et al.~\cite{kreibich2010netalyzr} initially used the measurement to determine whether home devices have Internet connectivity~\cite{agarwal2009netprints, dicioccio2012probe}. 
Prior works~\cite{grover2013peeking, sundaresan2014bismark} measure the availability, infrastructure, and usage of home networks.  
Shodan~\cite{Shodan} is the first search engine that continuously carry out network measurements from the whole IPv4 space. 
Several similar and follow-up search engines (Censys~\cite{Censys}, Zoomeye~\cite{Zoomeye}, Fofa~\cite{FoFA} and BinaryEdge~\cite{BinEdge}) appeared for collecting internet-connected devices.
All search engines rely on application banners to detect and profile online devices and services.
However, we have no idea what model/regex those search engines use.

Many prior works~\cite{Hershel2014, formby2016s, fachkha2017internet, miettinen2017iot, xuan2018are} leverage banners and other information to characterize IoT devices. 
Kumar et al.~\cite{kumar2019all} scanned IoT devices behind the NAT and provided the analysis on device security properties.
Huang et al.~\cite{huang2020iot} performed smart home device analysis by inspecting traffic management, where an ensemble of classifiers achieved high accuracy for the IoT device recognition.
Perdisci et al.~\cite{perdisci2020iotfinder} analyzed DNS application protocols to derive IoT model fingerprints by taking a document retrieval-based approach.
Saidi et al.~\cite{saidi2020haystack} leveraged behavior-based IoT device identification methods to reveal some information about the devices, e.g., the manufacturer.
Izhikevich et al.~\cite{izhikevich2021lzr} proposed an advanced scanning tool to identify Internet services and to reduce false positives due to middleboxes.
Different from prior works, we investigate the obfuscation technique against the IoT device scanning, in attempt to mislead attackers for their IoT information gathering.
{\tech} reveals that carefully crafted adversarial examples can easily deceive both the regex-based and learning-based scanners.

\textbf{Adversarial attack for obfuscation} is to craft or confuse the perturbations against machine learning models applied in many domains. 
In the OS fingerprinting domain, Smart et al.~\cite{smart2000defeating} was the first to limit an attacker's OS fingerprinting capabilities by removing relevant OS clues in TCP/IP packets. 
IP Personality~\cite{ip_personality} is a Linux kernel patch that can simulate other OS fingerprinting information at the TCP/IP layer, thereby fooling the fingerprinting tools. 
In the website fingerprinting domain, Abusnaina et al.~\cite{abusnaina2020dfd} used a random percentage of the total burst size in the traffic as the defense mechanism against a CNN model.
Ling et al.~\cite{ling2022towards} proposed a genetic-based variant to evade/obfuscate website fingerprints, where they injected dummy packets into the raw traffic as the defense strategy.
Mathews et al.~\cite{mathews2022sok} explored today's notable website fingerprinting defenses, where hand-crafted features may still leak information.
For the malware detection model, several prior works~\cite{grosse2017adversarial, yang2017malware} have investigated the applicability of adversarial attack techniques to malware classification.
Severi et al.~\cite{severi2021explanation} leveraged the backdoor attacks against machine learning models in malware classification, including Windows PE files, PDFs, and Android applications.
Different from prior works, we are the first to demonstrate adversarial examples of polluting automated IoT device scanners for acting as a feasible and promising pre-active defense mechanism.

\section{Conclusion}
\label{sec:con}

Device scanning is the first step in mounting a massive IoT-based cyberattack and user privacy protection. Our work explores a new defense mechanism called {\tech} to mitigate such IoT device scanning activities by generating adversarial examples. 
The key feature of {\tech} is to leverage an IoT-related semantic space and a visual similarity space to discover available manipulating perturbations for generating adversarial examples of IoT device banners.  
Our analysis further illuminates the fact that learning- and matching-based scanners are vulnerable to adversarial manipulations of their banners, providing useful guidelines for defense against similar network reconnaissance activities.

\bibliographystyle{IEEEtran}
\bibliography{ref/ref_ours, ref/ref_online, ref/ref_measurement, ref/ref_fingerprint,ref/ref_security, ref/ref_program, ref/ref_algorithm, ref/ref_advsecurity}


\appendix

\section{Appendix}

\begin{table}[hbt]
    \renewcommand{\arraystretch}{1.3}
    \caption{Region Partition for the HTTP banners. }
    \label{tab:region}
    \centering

    \begin{tabular}{ c c }
    \toprule
        {\bf Region}& {\bf Regex/Strings}\cr	
      \midrule
      {\bf IMR}
          &\makecell[l]{element tags in the HTML banner, CSS styles, \\Javascript in the
HTML content}\cr
      {\bf FR}
          &\makecell[l]{$<$h1$>$, $<$h2$>$, $<$h3$>$, $<$h4$>$, $<$h5$>$, $<$a$>$\\
                         $<$h6$>$, $<$B$>$, $<$big$>$, $<$em$>$, $<$i$>$,  $<$option$>$\\
                         $<$strong$>$, $<$title$>$, $<$dt$>$, $<$dd$>$, $<$li$>$ \\
                         $<$th$>$, $<$td$>$, $<$caption$>$ }\cr
      
      {\bf NFR}
          &\makecell[l]{$<$font$>$, $<$span$>$, $<$label$>$, $<$p$>$,  $<$!--$>$   \\ $<$blockquote$>$, $server$ in header,/* */ etc.}\cr 
    \bottomrule
    \end{tabular}
\end{table}

\begin{table}[hbt]
        \renewcommand{\arraystretch}{1.3}
        \caption{ Reversed order character set.}
        \label{tab:unicode}
        \centering
        \fontsize{7}{7}\selectfont
        \begin{tabular}{lll}
        \toprule
            {\bf }&{\bf Unicode}&{\bf Usage}\cr	
            \midrule
            {\bf LRO}&\makecell[c]{U+202D}
                    &\makecell[l]{Forcing the text to be displayed from left to right}\cr
            {\bf RLO}&\makecell[c]{U+202E}
                    &\makecell[l]{Forcing the text to be displayed from right to left}\cr
            {\bf PDF}&\makecell[c]{U+202C}
                    &\makecell[l]{Terminates the scope of action of the LRO and RLO \\ and restores the bidirectional state to its previous state.}\cr
            {\bf  LRI}&\makecell[c]{U+2066}
                    &\makecell[l]{Isolation of text from left to right.}\cr
            {\bf PDI}&\makecell[c]{U+2069}
                    &\makecell[l]{Terminates the scope of action of the LRI}\cr
            {\bf BS}&\makecell[c]{U+0008}
                    &\makecell[l]{Backspace.}\cr	
        \bottomrule
      \end{tabular}
\end{table}

\begin{figure}[hbt]
    \centering
    \includegraphics[width=3in]{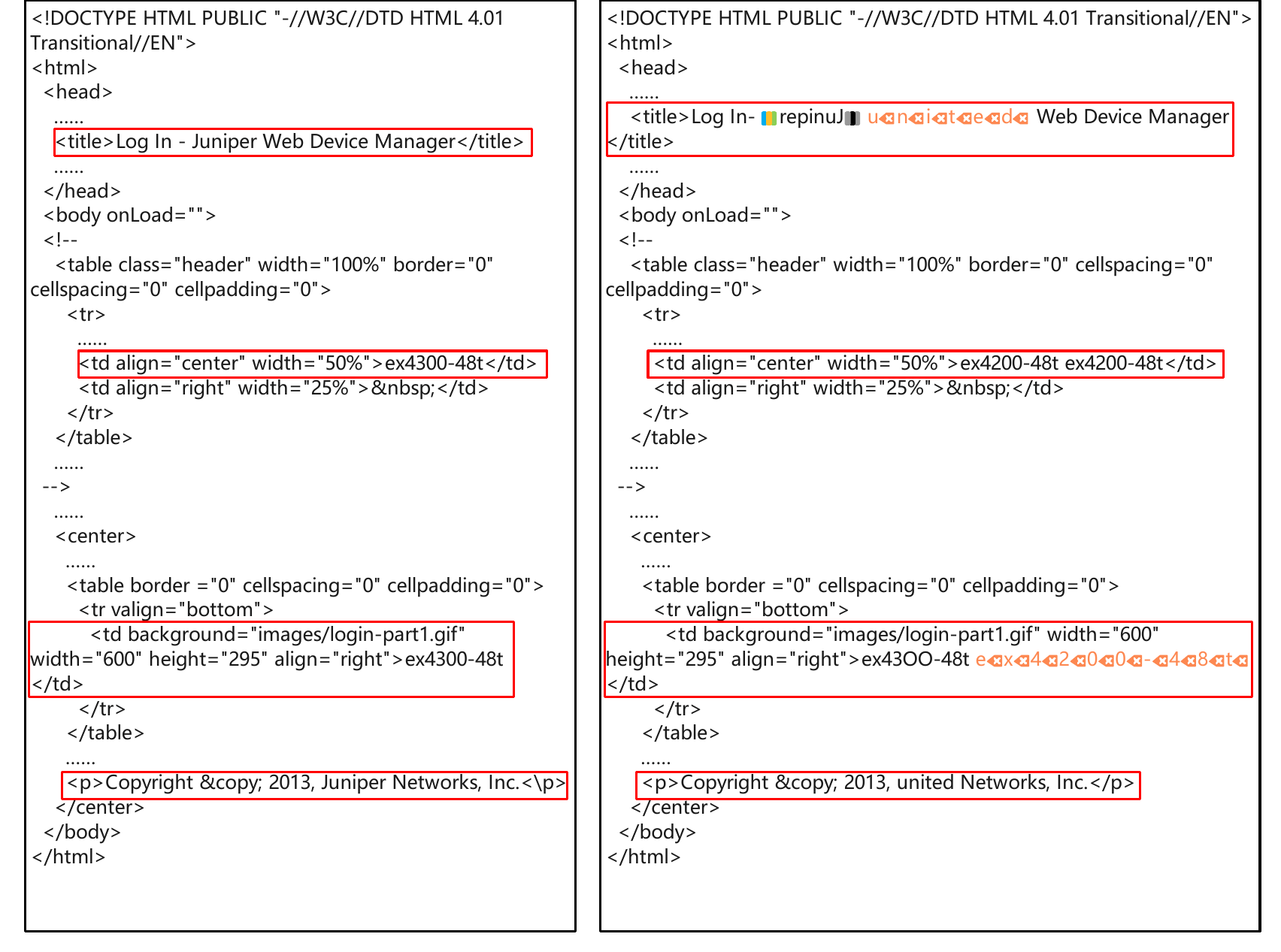}
    \caption{ The IoT banner (HTML format) with/without Unicode specifications. The left is the original input, and the right is the perturbed one, where the red box is the place perturbed by the unicode.}
    \label{fig:unicode:html}
\end{figure}

\begin{figure}[hbt]
    \centering
    \includegraphics[width=3 in]{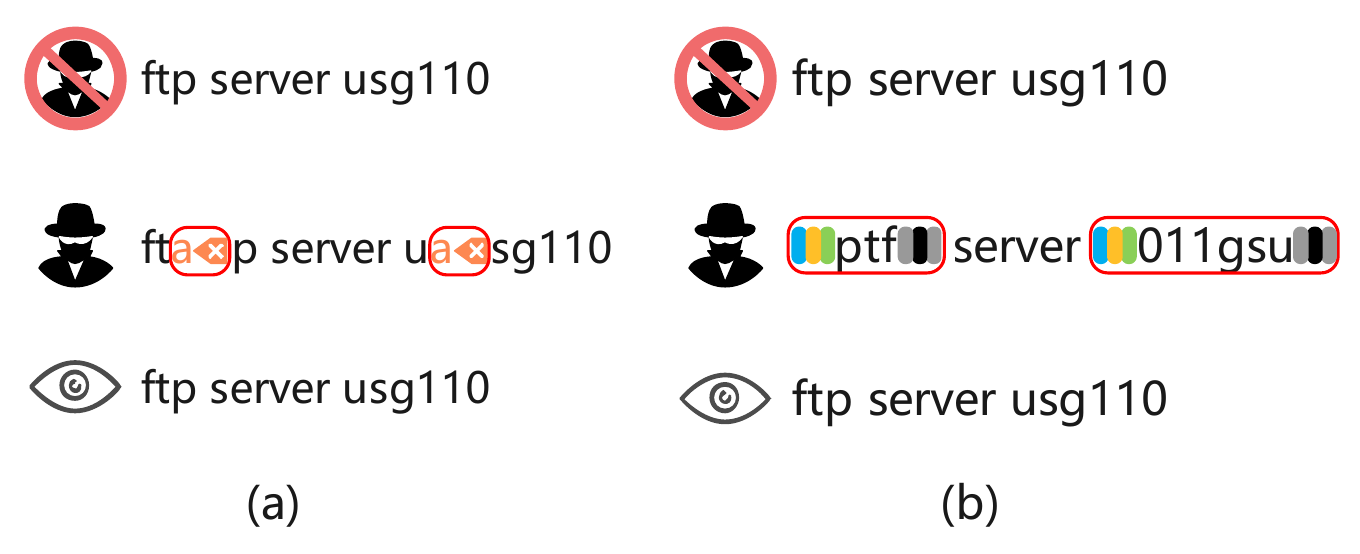}
    \caption{The effect of different Unicode or Unicode pairs.
    The part of (a) framed in red in is a combination of a character and BS (U+0008). The character before the BS is visually hidden.
    The part of (b) framed in red indicates that the string is encoded in reverse order, surrounded by three pairs of Unicode control characters. The blue block indicates LRO (U+202D), the yellow block indicates LRI (U+2066), the green block indicates RLO (U+202E), the grey block indicates PDF (U+202C) and the black block indicates PDI (U+2069). They need to be used in pairs and serve to display strings encoded in the reverse order into the correct order.}
    \label{fig:unicode}
 \end{figure}

\begin{table}[!t]
	\renewcommand{\arraystretch}{1.3}
	\caption{Parameters of the model for deriving IoT devices}
	\label{tab:model:parameter}
	\centering
	\fontsize{7}{7}\selectfont
	\begin{tabular}{  c c c c c c }
		\toprule
					
		\multirow{2}{*}{\bf Model}& \multicolumn{2}{c}{\bf Parameter} & \multirow{2}{*}{\bf Model}& \multicolumn{2}{c}{\bf Parameter}\cr
					
			\cmidrule(lr){2-3} \cmidrule(lr){5-6}  &Name &Value &   &Name &Value  \cr
							
				\midrule
					\multirow{5}{*}{\bf CNN}    &Units&256 &  \multirow{5}{*}{\bf LSTM} &Units&128,64 \cr
						  &Batch Size&128 &   &Batch Size&128  \cr
						  &Learning Rate&1e-4 & &Learning Rate&1e-3\cr
						  &Epochs&10,10,13 & &Epochs&3,6,16\cr
						  &Dropout&0.5  &  &Dropout&0.5\cr
				
					\bottomrule
				\end{tabular}
\end{table}

\end{document}